\documentclass[12pt, preprint]{aastex}
\usepackage {graphicx}
\def\deg{$^{\circ}$}
\def\um{${\rm \mu m}$}

\begin{document}

\title{HST/NICMOS Observations of Fast Infrared Flickering in the
Microquasar GRS 1915+105}

\author{Stephen S. Eikenberry\altaffilmark{1}, Shannon
G. Patel\altaffilmark{1}, David M. Rothstein\altaffilmark{2}, Ronald
Remillard\altaffilmark{3}, Guy G. Pooley\altaffilmark{4}, Edward
H. Morgan\altaffilmark{3}}

\altaffiltext{1}{Department of Astronomy, University of Florida, Gainesville, FL  32611}
\altaffiltext{2}{Department of Astronomy, Cornell University, Ithaca, NY 14853}
\altaffiltext{3}{Department of Physics, Massachusetts Institute of Technology, Cambridge, MA  02138}
\altaffiltext{4}{Cavendish Laboratory, Cambridge University}

\begin{abstract}
  
  We report infrared observations of the microquasar GRS 1915+105 using the
  NICMOS instrument of the Hubble Space Telescope during 9 visits in April-June
  2003.  During epochs of high X-ray/radio activity near the beginning and end
  of this period, we find that the $1.87 $\um  infrared flux is generally low
  ($\sim 2$ mJy) and relatively steady.  However, during the X-ray/radio
  ``plateau'' state between these epochs, we find that the infrared flux is
  significantly higher ($\sim 4-6$ mJy), and strongly variable.  In
  particular, we find events with amplitudes $\sim 20-30$\% occurring on
  timescales of $\sim 10-20$s (e-folding timescales of $\sim 30$s).  These
  flickering timescales are several times faster than any previously-observed
  infrared variability in GRS 1915+105 and the IR variations exceed
  corresponding X-ray variations at the same ($\sim 8s$) timescale. These
  results suggest an entirely new type of infrared variability from this
  object.  Based on the properties of this flickering, we conclude that it
  arises in the plateau-state jet outflow itself, at a distance $<2.5$ AU from
  the accretion disk.  We discuss the implications of this work and the
  potential of further flickering observations for understanding jet formation
  around black holes.

\end{abstract}
 
\keywords{ accretion, accretion disks --- black hole physics --- infrared: stars --- stars: individual (GRS 1915+105) --- X-rays: binaries }

\section{Introduction}

As the archetypal microquasar, GRS 1915+105 offers great promise as the
potential key for understanding the formation of relativistic jets in black
hole systems.  GRS 1915+105 was first discovered as a hard X-ray transient
\citep{alberto}, and then later seen as the first-known superluminal jet
source in the Galaxy \citep{mr94}.  Further study revealed a rich variety of
X-ray, infrared (IR), and radio variability from this system (see \citet{MRaa}
and references therein).  Among these, \citet{fp97} discovered radio and
infrared flaring on the same day (in disjoint observations), each displaying a
$\sim 30$-minute quasi-periodicity.  Later simultaneous observations in the
X-ray/IR and X-ray/IR/radio showed that these events tied together activity in
the inner accretion disk (seen in the X-ray) with synchrotron jet ejections
(seen in the IR and radio) (\citet{eiken98a}; \citet{felix98}; \citet{fp99}).
Thus, GRS 1915+105 seemed to be living up to its potential as a prototype
source for investigations of the disk-jet connection in black hole binaries.

Since then, continued study has revealed that the disk-jet connection in GRS
1915+105 takes multiple, apparently distinct, forms.  \citet{eiken00} describe
three types of long-wavelelength variability: large-amplitude superluminal
radio events (Class A); the $\sim 30$-minute IR/radio oscillations noted above
associated with hard X-ray dips (Class B); other IR/radio flares associated
with individual soft X-ray oscillations (\citet{eiken00}; \citet{rothstein04})
(Class C).  The Class B and C events share similar radio properties (e.g.
\citet{kleinwolt02}), but the Class B IR flares are significantly brighter
($\sim 200-300$ mJy) than their class C counterparts ($\sim 5-10$ mJy), and
the X-ray properties of these differ (\citet{rothstein04}; \citet{mikles}).
Both of the observed IR event types have e-folding rise times of $\sim
200-300$s.

To date, no IR counterparts to Class A events have been reported, though there
is some evidence for extended IR jet emission (\citet{sams}; \citet{eiken96}).
This has been due in part to the transient nature of these events, with
occurrence rates of only one every few years since 1994.  However, detailed
study has revealed an apparent pattern in these events which could allow the
prediction of Class A events with lead times of a few weeks.  Occasionally,
GRS 1915+105 enters a ``plateau state'' (\citet{Foster96}; \citet{Fender99}),
where the X-ray count rates are relatively low, dominated by a hard power-law
component, and with strong quasi-periodic oscillations at frequencies $\sim
1-10$ Hz.  The radio emission from GRS1915+105 is generally elevated and
quasi-steady ($\sim 50-200$ mJy).  It seems that the beginning and end of these
plateaux are marked by large-amplitude radio flares \citep{kleinwolt02}, and
that just before they post-plateau flare the X-ray hardness ratio begins
curving towards softer values over $\sim 10-20$ days.  By identifying plateaux
via X-ray and radio monitoring observations, we can thus hope to schedule
ground- and space-based telescope resources catch the IR behavior of GRS
1915+105 during these events.  With this goal in mind, we proposed Target of
Opportunity (TOO) observations of GRS 1915+105 using the Near-Infrared Camera
and Multi-Object Spectrograph (NICMOS) instrument \citep{NICMOS} on the Hubble
Space Telescope (HST), triggered by X-ray/radio observations of a plateau
state nearing its end, as observed by the All-Sky Monitor (ASM) of the Rossi
X-ray Timing Explorer (RXTE) and the Ryle Telescope.

In the sections below, we first describe the observations taken as
part of this program, and the data reduction techniques used to
analyze them.  We next describe the results of these analyses,
including the identification of fast IR flickering from GRS 1915+105
during the plateau state.  We then move on to discussion of the
implications of these observations for undertsanding the formation of
jets in GRS 1915+105.  Finally, we present the conclusions based on
these results.

\section{Observations and Data Analysis}

\subsection{Infrared Observations}

The NICMOS observations we report here were taken as TOO data with
HST.  We triggered the TOO on MJD 52732 (3 April 2003), based on the
combination of a ``plateau'' state in GRS 1915+105 lasting $> 10 $
days, and the beginning of a curving drop in the HR2 hardness ratio in
data from the RXTE ASM.  We then observed GRS 1915+105 using NICMOS
over the span of nine visits from 2003 April 7 to 2003 June 24 (see
Table 1).  For each visit, we used the F187W filter ($1.75-2.0 $\um)
with the NIC2 detector ($256 \times 256$-pixel HgCdTe array, $
19.2\arcsec\ \times 19.2\arcsec$ field of view) for imaging
observations and the F240M ($2.3-2.5 $\um) and G206 ($1.4-2.5 $\um)
direct-grism filter pair with NIC3 ($256 \times 256$-pixel HgCdTe
array) for spectroscopic observations.  We also made polarimetric
observations with the long-wavelength polarizers ($1.9-2.1 $\um) and
NIC2.  All exposures were set to MULTIACCUM mode, with multiple
non-destructive reads allowing us to analyze the behavior of the
source on time scales shorter than the total exposure times.  For each
of the first three visits, we obtained a set of 5 dithered 72-s images
using the F187W filter, and a set of 12 dithered 56-s polarized images (4
in each of 3 polarization angles: 0\deg, 120\deg, 240\deg).

We processed data taken with the F187W and polarization (POL) filters
through the standard NICMOS calibration pipeline, CALNICA.  We also
processed the grism data with CALNICA, with the exception of the
flatfield correction, which was determined upon extraction of
individual spectra through CALNICC.  We further processed and combined
all dithered images with CALNICB.

For each F187W MULTIACCUM frame, we carried out photometry for GRS 1915+105
and for a nearby comparison star (Star A -- see e.g. \citep{eiken98a}) in the
same field of view.  We defined an aperture with a 6 pixel radius
($\sim$0.5\arcsec) around each star, which just encompassed the first
diffraction ring, and subtracted the background flux from an annulus with
inner radius of 15 pixels ($\sim$1.1\arcsec) and outer radius of 18 pixels
($\sim$1.4\arcsec) to calculate the flux.  We found typical background count
rates for the F187W filter of $\sim 0.15\ {\rm DN/s}$.  We replaced any
isolated bad pixels within the photometric aperture with the median of the
adjacent pixels.  We omitted results for a star in a given readout if groups
of bad pixels were located within its aperture.  We calculated the uncertainty
in background-subtracted flux using the associated error array for each
exposure, and then multiplied by the PHOTFNU keyword in the header to convert
to units of flux density.  We calculated values of flux density for MULTIACCUM
exposures at NICMOS STEP8 time intervals by first extracting individual
readouts from the IMA FITS file extensions, and then determining the count
rates during a given interval by finding the difference in total counts
between readouts and dividing by the time interval.

We also included the polarimetry data in our photometric results using the
same calculation procedure as above for the F187W data.  (Due to
highly-variable IR emission from GRS 1915+105 during this epoch (see below)
and the sequential procedure of the polarimetry measurements with NICMOS, we
could not extract meaningful polarimetric measurements for the core source.)
We found background count rates for the POL filters of $\sim 0.23 \ {\rm
  DN/s}$.  While differential polarization may affect absolute flux
comparisons between GRS 1915+105 and Star A, it is only on the order of a few
percent or less, and should not significantly impact measurements of
time-variability in flux within a given filter/polarization.

\subsection{Radio Observations}

The Ryle Telescope has been used to monitor a small number of variable and
transient sources, including GRS 1915+105, at 15 GHz (during gaps in its primary
CMB-related program).  The observational details are as described by
\citet{pf97}; typical runs are for an hour or so, with longer times for some
coordinated programs.  The data points shown in Fig 1 are 5-min averages, on
which the noise level is about 2 mJy rms, and the overall flux-density scale
has an uncertainty of about 4\%.

\subsection{X-Ray Observations}

For X-ray observations of GRS 1915+105, we used both the All-Sky Monitor (ASM)
and the Proportional Counter Array (PCA) instruments of the {\it Rossi X-ray
  Timing Explorer} (RXTE).  The ASM uses scanning masked proportional counters
to provide ``snapshot'' observations of X-ray sources distributed across the
sky, with typical sampling frequencies of several measurements per day (for
details on the ASM, see \citet{lev96}).  For monitoring the X-ray behavior of
GRS 1915+105 during the IR observations, we used the dwell-by-dwell count rate
and hardness ratio (HR2) downloaded from the ASM web site.

The PCA instrument on RXTE uses up to 5 proportional counter units (PCU) with
large collecting area for detailed pointed observations (for details on the
PCA, see \citet{jah2006}) .  For specific pointed observations, we analyzed
the light curves, energy spectra, and power density spectra of GRS 1915+105.
We used the FTOOLS package for RXTE to generate light curves with time
resolutions of 1-second (using the Standard1 data mode) and 0.020-seconds
(using the binned 8ms data mode).  For spectral analyses, we used the XSPEC
package together with FTOOLS v5.2 to analyze Standard2 data (16-second time
resolution and 129 energy channels over the 2-60 keV energy range).  We used
standard procedures for response matrix generation, background estimation and
subtraction, and correction for PCA deadtime.  We then used XSPEC v11.2 to fit
the 2.9-25 keV spectrum with a standard model for black hole candidates
consisting of a ``soft'' component (which peaks in the low-energy X-rays''
modeled as a multi-temperature disk blackbody (e.g. \citet{mitsuda}) and a
``hard component'' (which extends to the higher-energy X-rays) modeled as a
power law, both modified by interstellar absorption corresponding to a fixed
hydrogen column density of $N_H = 6 \times 10^{22} {\rm cm^{-2}}$
\citep{muno99}.  We added a systematic error of 1\% to the spectrum before
performing the fit.

\section{Results}

\subsection{The Multi-Wavelength Context}

Figure 1 shows the multi-wavelength behavior of GRS 1915+105 during
the period of the Spring 2003 HST observations.  The first $\sim 25$
days of these lightcurves show GRS 1915+105 to be in a typical
``plateau'' state, with low, steady, hard X-ray emission and elevated
radio flux.  Near MJD 52735, the X-rays reveal a sharp outburst
accompanied by radio variability -- the curvature in X-ray hardness
ratio leading up to this caused us to trigger the HST TOO as described
above, in anticipation if a plateau-ending outburst.  However, rather
than showing a very large outburst signalling the end of the plateau,
GRS 1915+105 showed a rather modest outburst and then {\it returned}
to the plateau state.  It then continued on in this state until near
MJD 52785.  Then, a second outburst led to the actual termination of
the plateau state.

In Table 1, we present the average IR flux densities for each visit, star and
filter type (F187W and POL).  Note that the nine data points in Figure 1c
indicate the mean flux density value for each visit, while the bars represent
the {\it range} of flux density values rather than uncertainty.  The light
curve in Figure 2 displays the range of IR flux density for GRS 1915+105 and
also presents the mean flux density of the comparison star for each visit.  In
Figure 2, the error bars represent the standard deviation in the measured flux
of the comparison Star A over the nine visits, which also gives a
reasonable estimate of the uncertainty in flux density for GRS 1915+105.
These errors are significantly larger than those derived from the CALNICA
pipeline error arrays, indicating some additional uncertainty in the
photometry.  However, when we use the CALNICA error estimates to fit a
constant flux density to Star A {\it within} a single visit, we obtain reduced
chi-squared values of $\sim 1$, indicating that the CALNICA error estimates
are valid within a single visit.  We attribute the additional between-visits
variations in the Star A photometry to possible PSF variations from the
telescope and/or NICMOS instrument.  We can see from Figure 2 that GRS
1915+105 is higher in both flux and fractional variability in Visits 2-4, as
compared to Visit 1 and Visits 6-9.  Interestingly, these visits (2-4)
correspond to the time period {\it between} the major X-ray/radio variability
episodes described above.  When the X-ray flux is high/variable, the IR flux
density seems suppressed and relatively steady.

\subsection{Fast IR Flickering}

To further illustrate the intra-visit IR variability from GRS 1915+105, we
present selected segments of the light curves in Visits 2-4 in Figure 3, along
with the accompanying photometry of Star A for comparison.  At times, GRS
1915+105 exhibits significant ($>20 \%$ amplitude) variability from one
8-second MULTIACCUM readout to the next.  We note that the visit-to-visit
instability in the photometry is not an issue for these rapid fluctuations, as
shown by the stability of Star A, which has similar brightness to GRS
1915+105, but is much more stable over these same timespans.  As noted above,
the residuals of the flux density values for Star A to a best-fit constant
value have $\chi^2_{\nu} \sim 1$, indicating that these uncertainties are
accurate within a given visit.  Typical events in Figure 3 have $\sim 5-10
\sigma$ statistical significance.  Thus, we conclude that this flickering is
in fact intrinsic variability in GRS 1915+105.  We quantify the mean IR flux
at $1.87 $\um and the sample RMS excess variability in that band (8 s bins) in
Table 1, so that the IR flickering rates can be compared within the set of HST
results and also with respect to results from other wavebands.

We characterized the timescale of these flickering events assuming an
exponential rise time for IR flares.  We derived characteristic
e-folding timescales using adjacent MULTIACCUM readouts (Table 2),
calculating the uncertainty by simple propagation of errors from the
flux density uncertainties.  The upper-limit timescale column in Table
2 indicates the maximum characteristic rise time based on the highest
possible flux value (within uncertainties) at the beginning of a time
interval and the lowest possible flux value at the end of it.  Many
flickering events show best-estimate e-fold timescales of $\sim 30$s,
and several have firm upper limits on this number of $<40$s.  We note
that these timescales are nearly an order of magnitude shorter than
the fastest previously-reported timescales for long-wavelength
(IR/radio) variability from GRS 1915+105 (e.g. \citet{eiken98a};
\citet{eiken00}).

\subsection{X-ray Behavior Associated with the Fast IR Flickering}

We have examined the X-ray spectra and power density spectra (PDS)
derived from RXTE pointed observations of GRS~1915+105 obtained near
the times of the HST observations (Table 1).  Particularly relevant
are the RXTE observations made on 5 days (2003 April 10, 16, 23, May 1
and 7) that are interleaved with HST visits 2-5.  The X-ray spectral
properties during this epoch of rapid IR flickering are typical of the
radio plateau state, showing a power-law dominated spectrum with a
photon index, $\Gamma \sim$ 2.1--2.3 (see \citet{Muno01}).  The PDS
during these times are similar, and they are likewise typical of the
plateau state.  The first two of these PDS (April 16 = MJD 52745.1 and
April 23 = MJD 52752.2) are shown in Fig. 4, using the full PCA
bandpass, which is effectively 2--40 keV.  These PDS are subtracted
for deadtime-corrected Poisson noise, normalized to power density
($P_{\nu}$) in units of $(rms/mean)^2$ Hz$^{-1}$, and then displayed
as log $(\nu \times P_{\nu}$ vs. log $\nu$.

The PDS shows a strong QPO just below 1 Hz, along with its second
harmonic.  These features are superposed on a strong and extremely
broad peak in the power continuum that is centered just above 1 Hz.
It has been argued that the broad power peak is a specific yet
unexplained X-ray signature of the steady jet in the hard state of
black hole binaries \citep{ron04}.  Since the X-ray PDS exhibits
little change over the interval of 2003 April-May, we can investigate
the X-ray and IR connection by comparing the flickering amplitudes at
the 8 s timescale of the IR light curve.  When the X-ray light curves
are examined in 8 s bins, the sample standard deviations for each of
the 5 relevant PCA observations are all in the range of 2.4--3.0 \% of
the mean count rate (2-40 keV).  This result represents the source
flickering amplitude, since the Poisson variations for these date are
well below 1\%.  These 8 s flickering amplitudes also represent the
X-ray behavior associated with the jet, since, as noted above, the
X-ray spectrum is dominated by nonthermal emission that is typical of
the plateau state, or more generally the X-ray hard state associated
with the presence of a steady radio jet \citep{jeffron06}.  One can
derive the analogous 8 s flickering amplitudes for the IR band using
the results given in Table 1.  The source RMS excess variability
amplitude (i.e. $\sqrt{\sigma_{sample}^2 - \sigma_{meas}^2} / mean$,
where $\sigma_{meas}$ are the measurement errors) is in the range of
4-8 \% for HST visits 2--5.  This exceeds the equivalent flickering
amplitude in the X-ray band by a factor of 2--3.  In addition, the
baseline IR flux contributed from the donor star could be in the range
of $\sim$ 2 mJy, in which case the jet flickering at 8 s would be
elevated to the range of $\sim 10$ \%

\section{Discussion}

\subsection{The Diversity of IR Variability in GRS 1915+105}

The type of rapid IR variability described above is particularly intriguing in
that it seems to be unrelated to the previous types of IR variability observed
in GRS 1915+105 (e.g. \citet{rothstein04}; \citet{eiken00}; \citet{reba}).
First of all, this IR flickering, while small in amplitude, is significantly
faster than previous IR flares, which seem to have e-folding timescales of
$\sim 200-300$s, as compared to the $\sim 25-60$s timescales seen here.  Thus,
the flickering is $\sim 4-10$ times faster than any previously-reported IR
variability from GRS 1915+105.  Second, in these observations we see no
apparent association between IR flickering events and any similar X-ray
events.  This is also in stark contrast to previous IR flares of various types
in GRS 1915+105, all of which seem to be associated with significant activity
in the X-ray band as revealed through relatively large-amplitude changes in
the X-ray lightcurve and/or spectrum (e.g. \citet{rothstein04};
\citet{eiken00}; \citet{eiken98a}; \citet{eiken98b}; \citet{felix98};
\citet{fp97}).  We note that our observations above {\it do} reveal
significant X-ray variability on {\it faster} timescales than the IR
flickering.  However, any X-ray variability is much smaller in amplitude when
smoothed to the same time resolution as the IR data.  Thus, the rapid IR
flickering we report here seems to be an entirely new kind of IR variability
in GRS 1915+105.

\subsection{IR Emission and Relativistic Jets in the Plateau State}

As shown above (Figures 1-2), the IR flux density and fractional variability
from GRS 1915+105 are significantly enhanced during the plateau state, even
though the X-ray flux is lowest and steadiest during this time period.  A
number of observations indicate that during such plateaux, GRS 1915+105
exhibits a quasi-continuous radio synchrotron-emitting relativistic jet
outflow which also disappears after the plateau-ending X-ray/radio outburst
(see \citet{kleinwolt02} for a detailed discussion).  As also shown above, the
IR flickering does not seem to correlate well with any particular X-ray
variability from GRS 1915+105.  The typical timescales for X-ray variability
(e.g. the $\sim 1$ Hz QPO) are much less than the 8-second time resolution of
the IR observations, and there is very little X-ray variability on these
longer timescales -- when smoothed to 8-second resolution, the X-ray
lightcurve has an RMS of $\sim 3$\% about a constant value.  Thus, while there
is significant activity in the inner accretion disk in GRS 1915+105 at this
time, it is rather weak on the timescales seen in the IR.  We conclude that it
seems unlikely that the IR flickering arises from the accretion disk which is
producing the jet.  Rather, based on this apparent simultaneity of the plateau
jet and the IR flickering, along with the lack of IR/X-ray correlation, it
seems sensible to conclude that the enhanced IR emission of a few mJy arises
from the quasi-continuous jet itself.  Previous authors have also suggested
that emission from the jet may enhance IR emission during the plateau state
(\citet{ueda02} ; \citet{fuchs} ).

Radio observations of this plateau jet indicate that the radio synchrotron
emission begins in the optically thick regime, and the optical depth in the
radio decreases to $\sim 1$ in the expanding jet at distances $\sim 50$ AU
from the accretion disk and compact object (\citet{kleinwolt02};
\citet{vivek}).  If we assume that the opening angle of this jet is comparable
to or larger than the large-scale jets observed from GRS 1915+105, then the
diameter of the jet is $\ge 1$ AU, or equivalently $\sim 500$ lt-s, at this
distance.  Given that the IR flickering occurs on timescales as fast as $\sim
25$s (20 times faster than the light travel time across the jet diameter at $d
\sim 50$ AU), we are forced to conclude that the IR flickering occurs much
closer to the source of the jet itself -- at a distance $\le 2.5$ AU.

We also note that the IR flux density excess seen here is $\sim 3-5$ mJy
(compared to observations outside the plateau state), corresponding to $\sim
100-200$ mJy once we correct for the extinction towards GRS 1915+105
(\citet{fp97}).  Assuming a distance of $\sim 12$ kpc for GRS 1915+105
\citep{mr94}, this gives an excess IR luminosity from the jet of $\nu L_{\nu}
\sim 10^{36} \ {\rm ergs/s}$.  This is much larger than the same number for
the radio flux from the plateau jet ($\nu L_{\nu} \sim 10^{32} \ {\rm
  ergs/s}$), indicating that the IR jet luminosity dominates the radio jet
luminosity by several orders of magnitude.  The overall luminosity of the
system at this stage, however, remains dominated by the X-ray flux from the
inner accretion disks, at levels of $\sim 10^{38} \ {\rm ergs/s}$.  Thus, the
radiated power in the jet we observe here is $\sim 1$\% of the total
luminosity of GRS 1915+105.

\subsection{What Drives the Fast IR Flickering?}

If the rapid IR flickering, unlike the previously-reported Class B and Class C
flares, does {\it not} arise in the inner accretion disk but the jet, we must
consider the question, ``What drives the rapid IR flickering on $\sim 30-60$s
timescales?''.  Without the direct influence of the inner accretion disk, the
most obvious possibility is that the timescales seen in the IR flickering are
the natural timescales of some sort of instability in the jet itself, near its
base.  Several effects near the jet base are expected to generate
shocks/instabilities.  First of all, the actual jet base -- the location where
particles are initially accelerated away from the compact object -- is
expected to be a significant source of non-thermal radiation.  While it is
generally considered that the non-thermal X-ray emission from these systems
(which is essentially {\it all} of the X-ray emission from GRS 1915+105 in the
plateau state) may be the primary radiation from the particle acceleration
process, it is not impossible that significant IR radiation arises here as
well.  Secondly (and perhaps more likely), there may be a jet collimation
point, where the previously-accelerated outflow from the jet base is
constrained to move with much smaller opening angles ($\sim 1$\deg).  Such
behavior is predicted by several theoretical models for jet formation, and
apparently confirmed by the radio observations of the M87 jet near its base
\citep{junor}.  If the jet base emission is in fact dominated by the
non-thermal X-rays, the particles may have ``cooled'' enough at the
collimation point to emit predominantly in the IR waveband.  For either of
these scenarios, further support may come from the fact that several authors
have previously identified an ``IR break'' in the spectrum of the X-ray binary
jet source GX 339-4 (\citet{motch}; \citet{CorbelFender}; \citet{nowak}).  In
some models of jet formation this break is directly associated to the
shock-acceleration region of the jet \citep{markoff} -- applying this to GRS
1915+105 would naturally lead to the interpretation of the flickering observed
here as coming from the jet launch region.  Finally, jet outflows from
supermassive black holes show clear evidence of internal shocks within the
collimated jet flow even further ``downstream'' from the collimation point.
If such shocks occur close enough to the compact object, they may also produce
rapid IR flickering as seen here.  Future observations of this flickering
which could further constrain these possibilities should give key insights
into the formation of the plateau jet in GRS 1915+105.

\section{Conclusion}

We have presented near-infrared observations of the microquasar GRS 1915+105
using the NICMOS instrument of the Hubble Space Telescope before, during, and
after a ``plateau'' state in March-May 2003.  During the X-ray/radio plateau,
we find that the infrared flux is significantly higher and more strongly
variable than outside the plateau.  In particular, we find ``flickering''
events with amplitudes $> 20$\% occurring on e-folding timescales of $\sim
30$s.  These timescales are several times faster than any previously-observed
infrared variability in GRS 1915+105, and do not seem to be correlated with
any X-ray events of similar amplitude/timescale.  As such, they seem to
represent an entirely new mode of IR variability from GRS 1915+105.  Based on
the simultaneity of this flickering with the plateau jet and the lack of
IR/X-ray correlation on the flickering timescales, we conclude that the IR
flickering arises in the plateau jet itself.  If so, the IR jet luminosity is
much greater than the corresponding radio jet luminosity.  Also, the IR
flickering must occur very near the base of the jet ($<2.5$ AU from the
accretion disk).  The lack of correlation with inner disk activity implies
that $\sim 30-60$s is a natural timescale of the jet at this range of
locations, and that the flickering may be due to shocks at either the site of
particle acceleration, the site of jet collimation, or internal shocks in the
jet.  Further study could give important insights into the formation of the
plateau jet in GRS 1915+105.

\acknowledgements {The authors wish to thank the staff of the Space
  Telescope Science Institute, particularly, N. Schulz, for their help and
  support for this TOO program with NICMOS.  The authors also thank the
  anonymous referee for helpful suggestions.  SSE and SGP were supported in
  part by a STScI grant, by an NSF CAREER award (AST-0328522), and NSF
  grant AST-0507547.  DMR was supported in part by a National Science
  Foundation Graduate Research Fellowship}

\begin{deluxetable}{lccc}
\tablewidth{0pt}
\tablehead{
\colhead{Visit}&\colhead{Date}&\colhead{F187W mean flux density}&\colhead{Excess Variability}\\
\colhead{}&\colhead{(MJD)}&\colhead{(mJy)}&\colhead{(\% RMS)\tablenotemark{a}} }
\startdata
1& 52736.60290 & 2.54 & 1.0 \\
2& 52741.67225 & 6.28 & 6.9 \\
3& 52746.87292 & 4.24 & 4.8 \\
4& 52752.87847 & 5.10 & 7.8 \\
5& 52763.68303 & 3.66 & 4.0 \\
6& 52796.43359 & 2.41 & 0.8 \\
7& 52801.71115 & 2.11 & $<2.0$ \\
8& 52808.31191 & 2.26 & 1.8 \\
9& 52814.71761 & 2.31 & $<2.0$ \\
\enddata

\tablenotetext{a}{Excess variability RMS \% is defined as
$\sqrt{\sigma_{sample}^2 - \sigma_{meas}^2} / mean$ where
$\sigma_{sample}$ is the RMS scatter in the sample values,
$\sigma_{meas}$ is the pipeline-estimated measurement noise RMS, and
$mean$ is the mean value of the sample points.  Values reported as
$<2.0$ \% inidcate that the RMS sample scatter is less than the
estimated RMS measurement noise, and we take $\sigma_{meas}$ as an
upper limit on $\sigma_{sample}$.}
\end{deluxetable}

\begin{deluxetable}{cccccc}
\tablewidth{0pt}
\tablehead{
\colhead{Visit}&\colhead{Filter}&\colhead{Date}&\colhead{UT}&\colhead{Characteristic}&\colhead{Upper}\\
\colhead{}&\colhead{}&\colhead{(MJD)}&\colhead{(s)}&\colhead{Rise Time (s)}&\colhead{Limit (s)}
}
\startdata
2 & F187W & 52741 & 58234 & $58\pm8$ & 71 \\
2 & POL0L & 52741 & 59079 & $31\pm4$ & 38 \\
2 & POL120L & 52741 & 59483 & $43\pm7$ & 55 \\
3 & F187W & 52746 & 75663 & $27\pm9$  & 50 \\
3 & F187W & 52746 & 75691 & $49\pm7$  & 61 \\
3 & F187W & 52746 & 75818 & $64\pm13$  & 88 \\
3 & POL120L & 52746 & 76755 & $52\pm9$  & 70 \\
4 & F187W & 52752 & 75904 & $28\pm9$  & 48 \\
4 & F187W & 52752 & 76289 & $45\pm5$  & 54 \\
4 & F187W & 52752 & 76392 & $55\pm6$  & 65 \\
4 & POL0L & 52752 & 76529 & $51\pm9$  & 69 \\
4 & POL0L & 52752 & 76735 & $48\pm9$  & 65 \\
4 & POL120L & 52752 & 77163 & $30\pm4$  & 37 \\
4 & POL240L & 52752 & 77369 & $28\pm3$  & 34 \\
4 & POL240L & 52752 & 77654 & $37\pm5$  & 45 \\
\enddata
\end{deluxetable}

\begin{figure}
\plotone{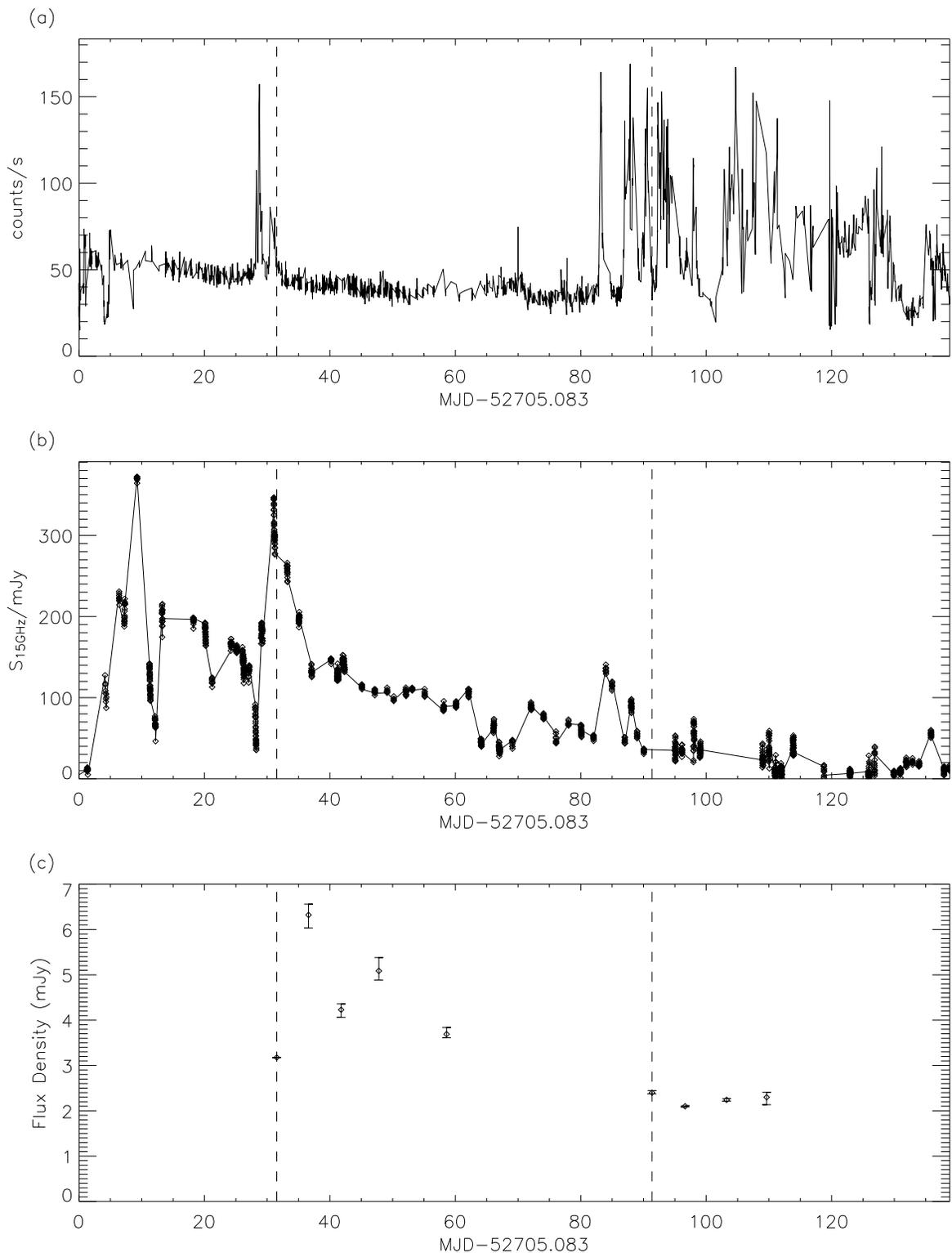}

\caption{Multiwavelength light curves of GRS 1915+105.  (Top) X-ray
  lightcurve from the RXTE ASM.  (Middle) Radio lightcurve from the Ryle
  Telescope.  (Bottom) Near-infrared ($1.87 $\um) lightcurve from
  $HST/NICMOS$.  For the near-IR lightcurve, note that the bars indicate the
  {\it range} of the variable flux density values during each observation,
  rather then the uncertainties.  Vertical lines provide a time reference for
  comparison among the different bands.}

\end{figure}

\begin{figure}
\plotone{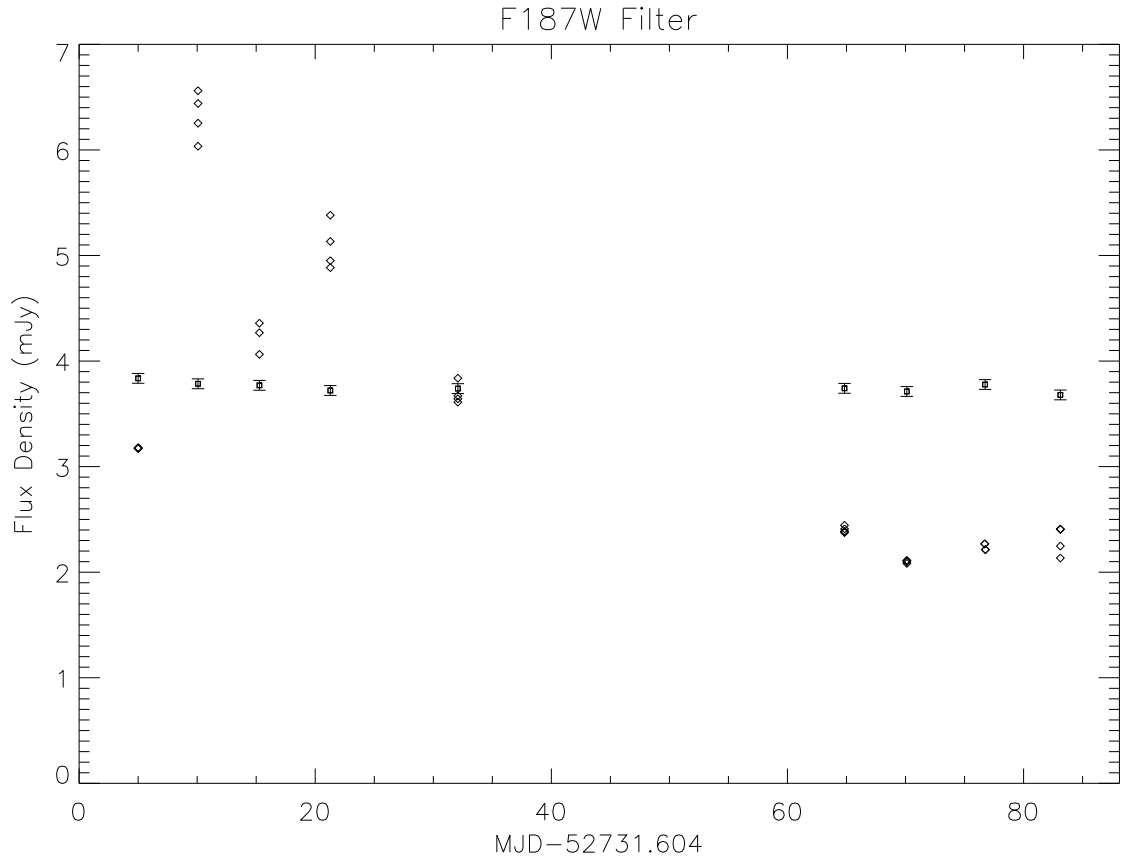}

\caption{Flux densities for GRS 1915+105 (Diamonds) and Star A
(Squares) in NICMOS F187W filter.  Note that Star A, which has similar
brightness to GRS 1915+105, is quite steady while GRS 1915+105 varies
both between visits and within individual visits.}

\end{figure}

\begin{figure}
\plottwo{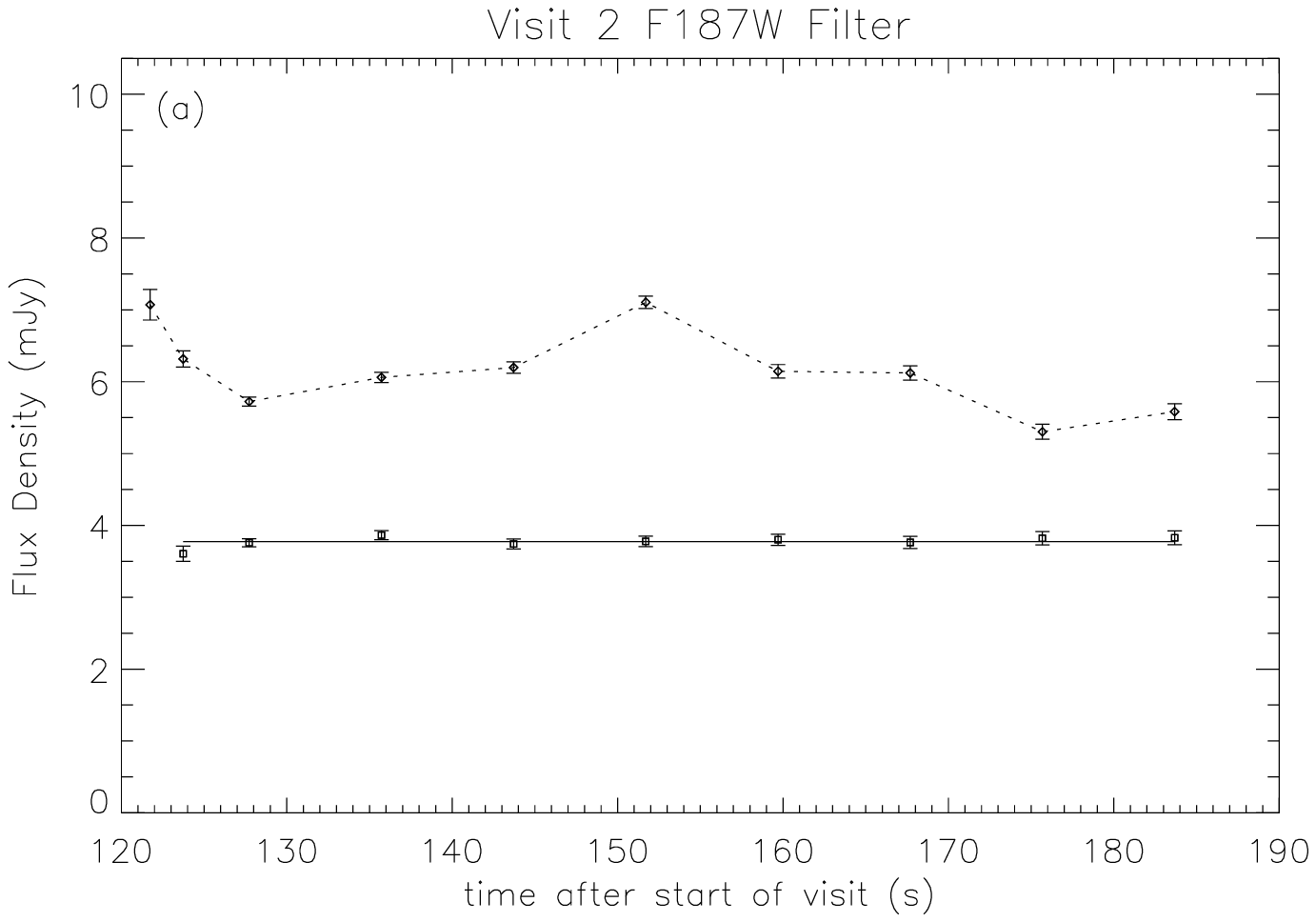}{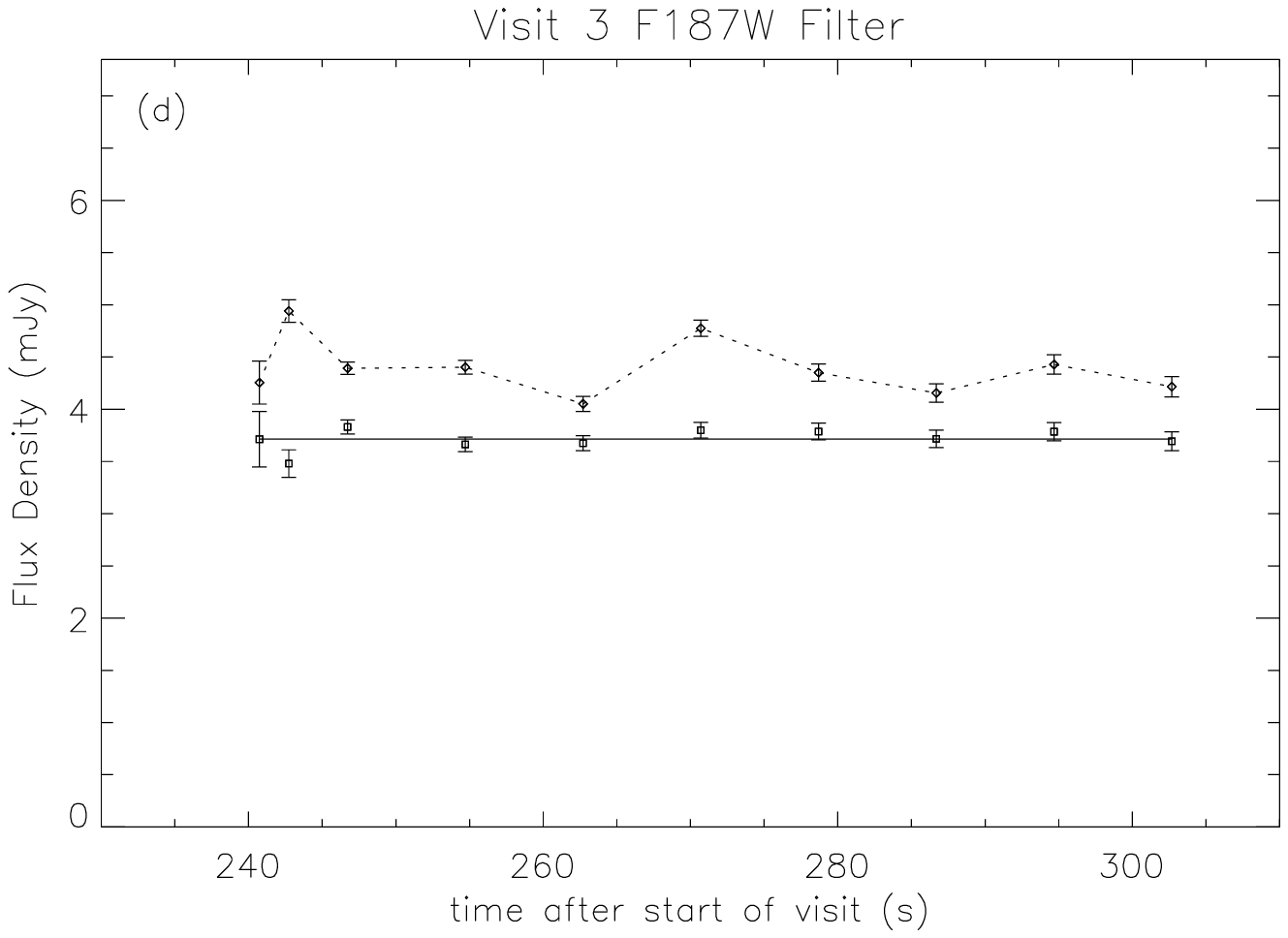}
\plottwo{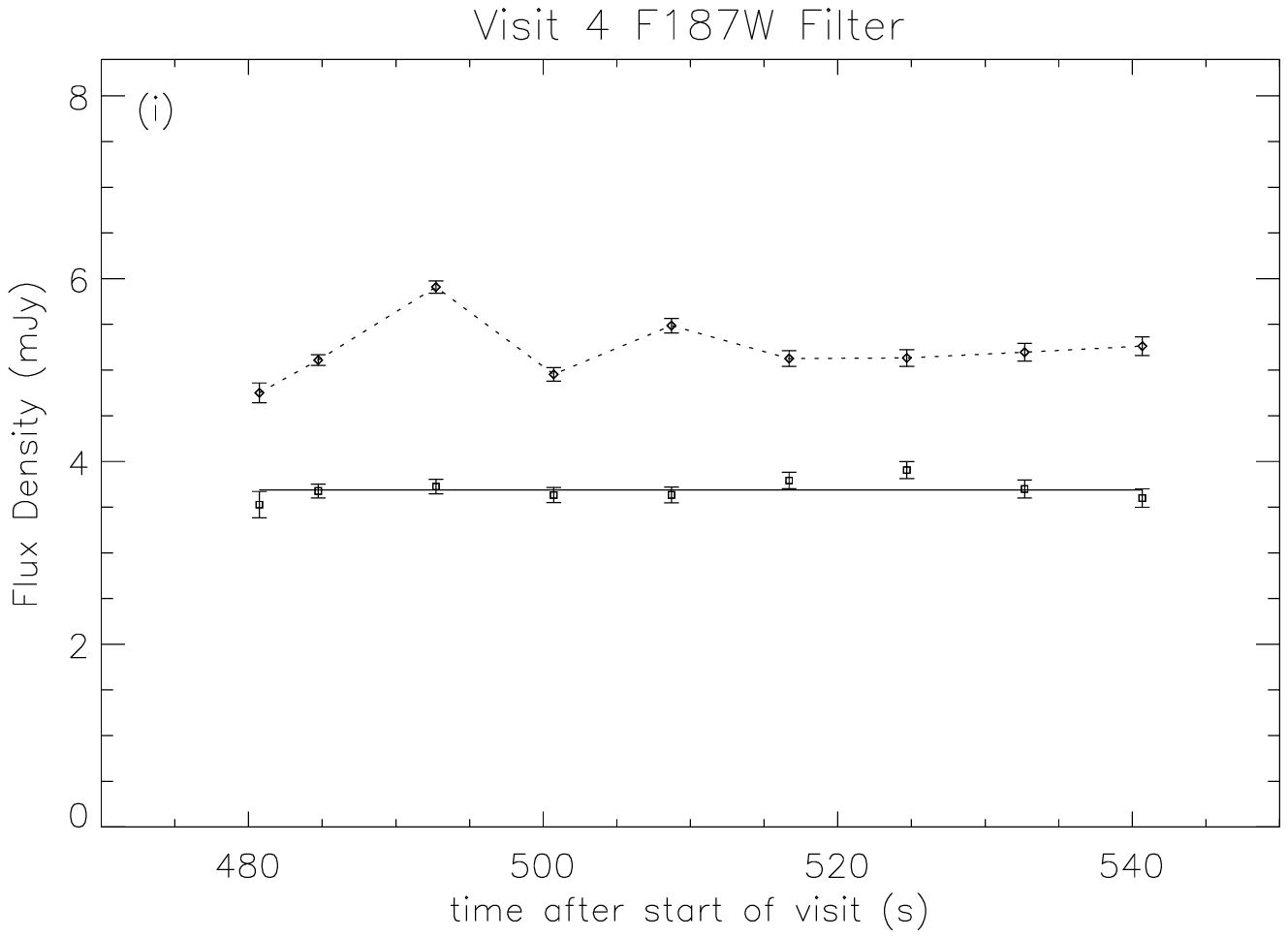}{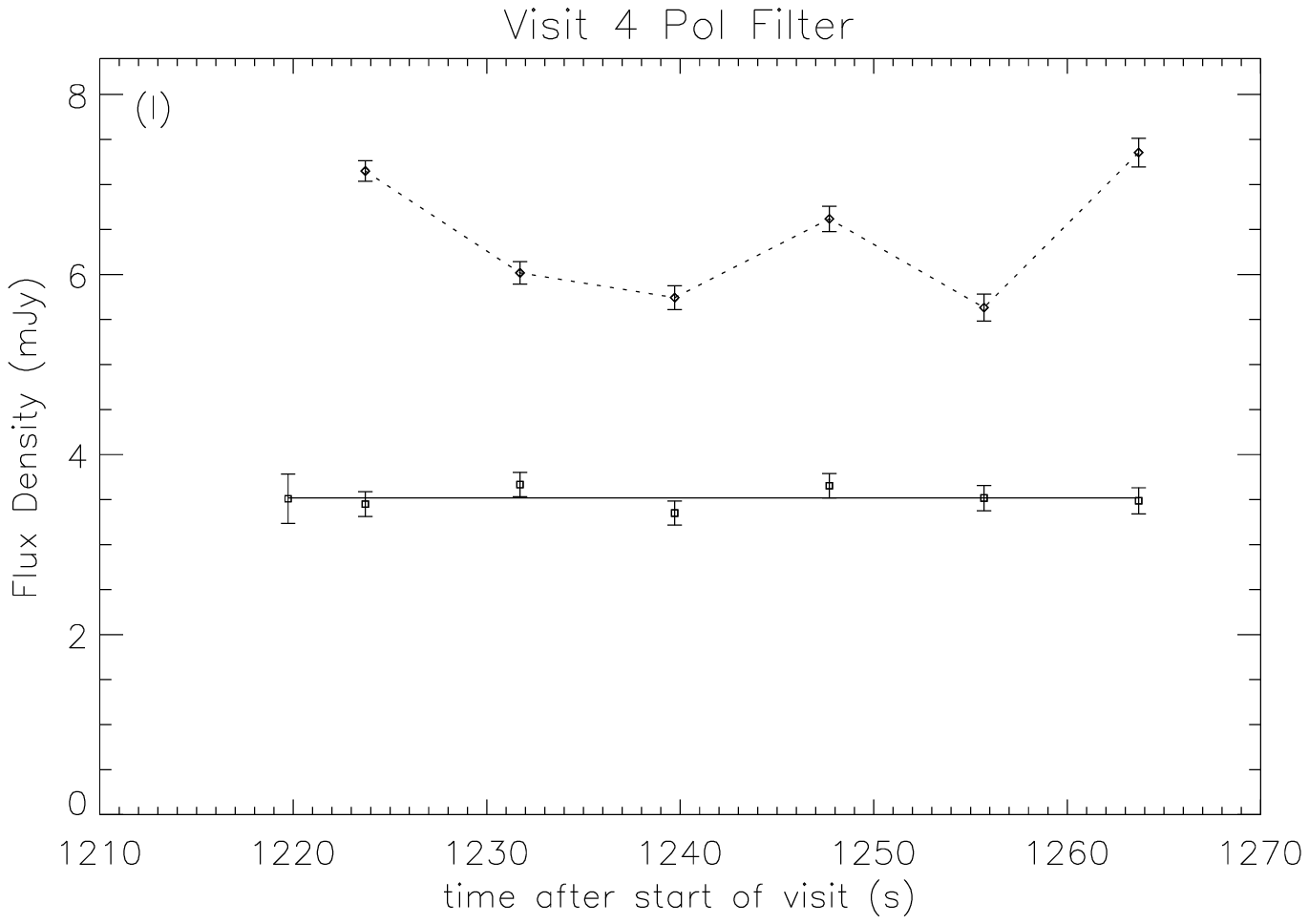}
\plottwo{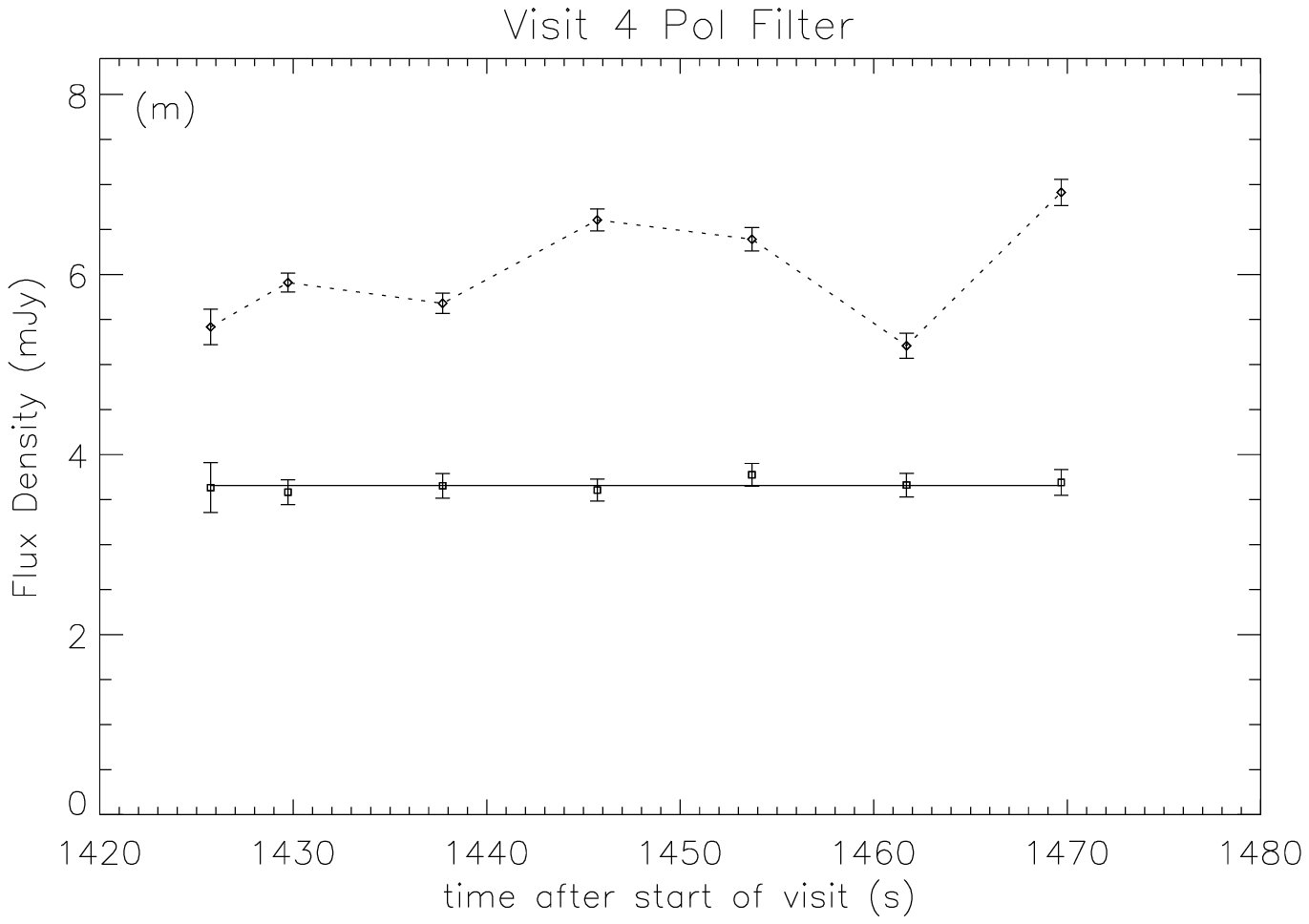}{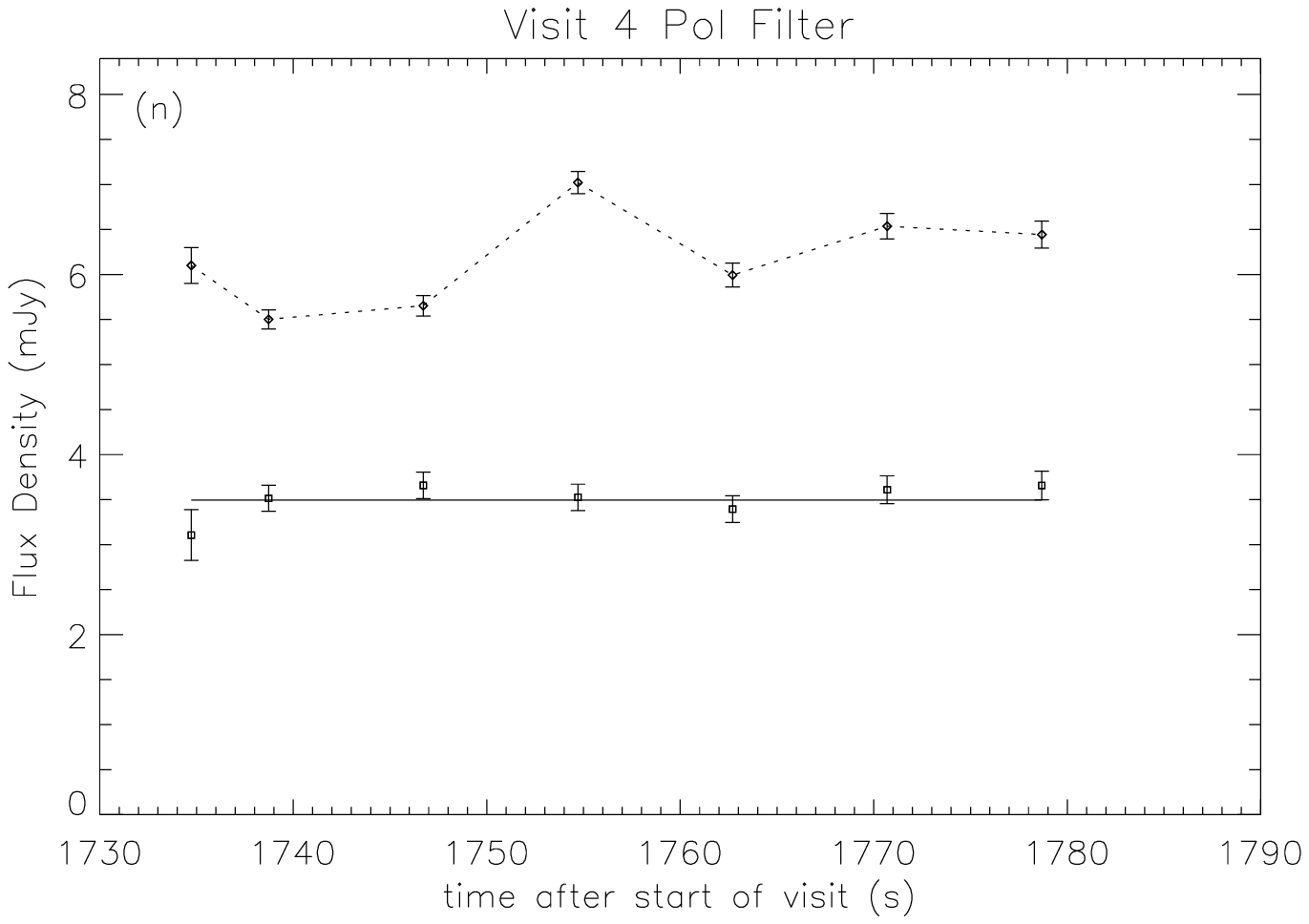}

\caption{Selected high time-resolution near-IR lightcurves for GRS
1915+105.}

\end{figure}

\begin{figure}
\plotone{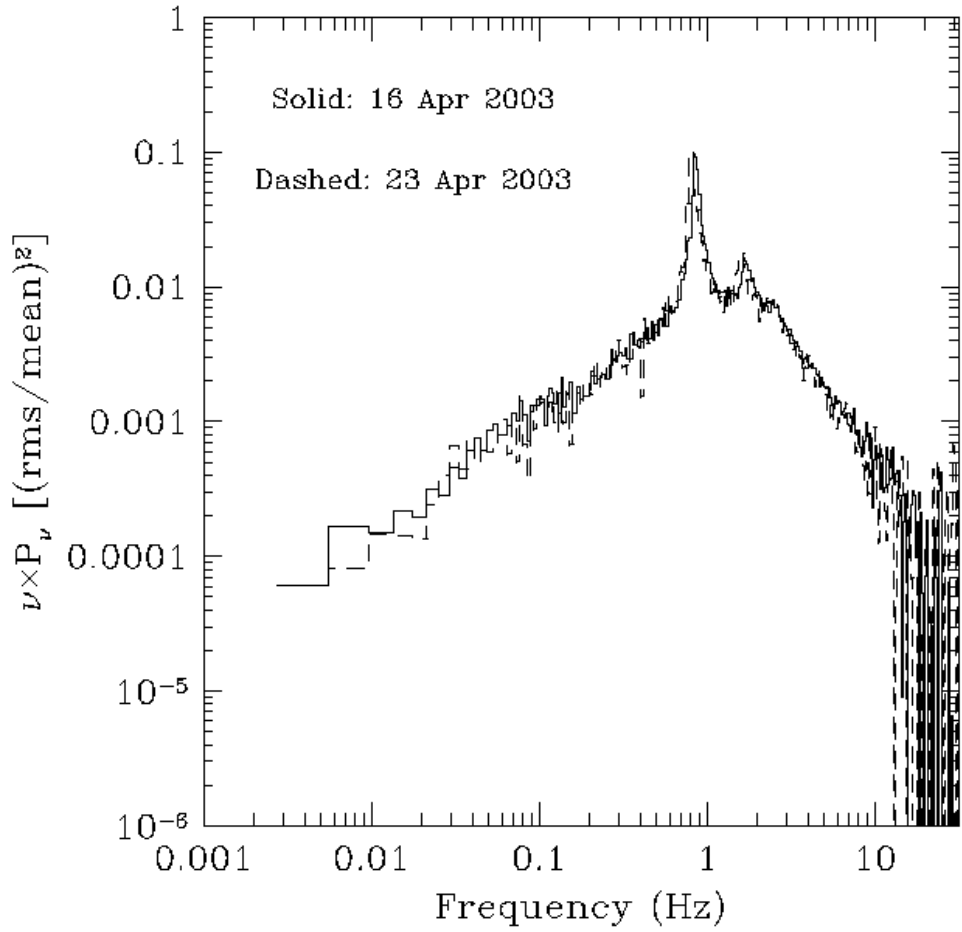}

\caption{X-ray power density spectra of GRS 1915+105 for 16 April 2003
(solid) and 23 April 2003 (dash).}

\end{figure}

\end{document}